\documentclass[seceq,preprint]{ptptex}

\usepackage{graphicx}

\catcode`\@=11
\def\lsim{\mathrel{\mathpalette\@versim<}}
\def\gsim{\mathrel{\mathpalette\@versim>}}
\def\@versim#1#2{\vcenter{\offinterlineskip
\ialign{$\m@th#1\hfil##\hfil$\crcr#2\crcr\sim\crcr } }}
\catcode`\@=12

\preprintnumber[4cm]{KANAZAWA-03-25}

\title{
Maximal Locality and Predictive Power
in Higher-Dimensional, Compactified Field Theories
}

\author{
Jisuke \textsc{Kubo} and Masanori \textsc{Nunami}
}

\inst{
Institute for Theoretical Physics, Kanazawa  University, \\
Kanazawa 920-1192, Japan
}

\abst{
To achieve a maximal locality in a trivial field theory,
we maximize the ultraviolet cutoff of the theory
by fine tuning the infrared values of
the parameters.
This optimization procedure is applied
to the scalar theory in $D+1$ dimensions ($D \geq 4$)  with
one extra dimension compactified on a circle with radius
$R$. The optimized, infrared values of
the parameters are then compared with the
corresponding ones
of  the uncompactified theory 
in $D$ dimensions, which is assumed to be
the low-energy effective theory.
We find that these values approximately agree 
with each other, as long as
$R^{-1} \gsim s M$ is
satisfied, where $s\simeq 10,50,50, 100$  for $D=4,5,6,7$, 
and $M$ is a typical scale
of the $D$-dimensional theory.
This result supports the previously made
claim
that the maximization 
of the ultraviolet cutoff in 
an nonrenormalizable field theory
can give the theory more predictive power.
}

\begin{document}

\maketitle

\section{Introduction}
Since Kaluza and Klein \cite{kaluza} found a possibility of 
unifying
fundamental forces by introducing extra dimensions, their idea has 
attracted attentions for many decades. Recently, there have been 
renewed interests in field theories  with extra dimensions.
\cite{Antoniadis}\tocite{dienes1} 
Since field theories in more than four dimensions
are usually nonrenormalizable,
the dependence of the UV cutoff can not
be completely removed, and moreover one has to introduce
infinitely many independent parameters in these theories.
In short, nonrenormalizable theories have much less
predictive power compared with renormalizable theories.
In our previous paper \cite{Kubo:2001tr}, we 
proposed a method, called maximal locality method,
to make nonrenormalizable theories
more predictive. 
The method is based on a simple intuitive picture
that the renormalization group (RG) flow of the effective theory
of a fundamental theory,
which evolves for ``maximal running time'', will be the best 
approximation to the renormalized trajectory of the fundamental theory.

In this work, we apply the method to compactified higher dimensional 
theories, and will consider 
in particular the  scalar theory in $D+1$ dimensions
with one dimension compactified on a circle with radius $R$.
One intuitively expects that
the  $D$ dimensional 
theory with all the Kaluza-Klein massive modes 
suppressed is the low-energy effective theory of the original
$(D+1)$-dimensional theory.
So, 
the predictions of the $D$ dimensional effective theory and the
original $(D+1)$-dimensional theory
should agree with each other at low energies.
Therefore, if maximal locality method is a sensible method,
it has to satisfy this consistency.
We  find that
there exits a maximal radius above which 
the consistency requirement can not be satisfied.

This paper is organized as follows. In Sect.~\ref{ML-method}, we
outline of the basic idea of maximal locality method with a concrete
example. 
In Sect.~\ref{Appli} we derive
a RG equation in the local potential approximation 
(the Wegner-Houghton equation) for compactified scalar theories.
The five-, six-,
seven-, and eight-dimensional scalar theories 
are considered in Sect.~\ref{results},
and we estimate the maximal radius for each case.
Lastly we conclude in Sect.~\ref{conclusion},
and the explicit expressions for the $\beta$-functions which we use in 
this paper are given in
Appendix\ref{KK-beta}.

\section{ Maximal Locality Method}
\label{ML-method}
\subsection{Formulation}
\label{formulation}

The basic idea of maximal locality method is based on a simple  intuitive picture.
Consider a  theory,  like QCD,  which
is free from the
UV cutoff, $\Lambda_0$, and suppose
its low-energy effective theory,
like non-linear sigma model, is
a perturbatively nonrenormalizable,
trivial theory that becomes weakly coupling in the infrared (IR) regime.
We then formulate both theories
in terms of the Wilsonian renormalization group (RG) \cite{wilson1}. 
Since we assume that the high-energy theory is free from the
UV cutoff, $\Lambda_0$,
we can let go $\Lambda_0$ to infinity.
In other words,
the RG flow  in the high-energy theory evolves along 
a renormalized trajectory, and
approaches an UV fixed point in the UV limit.
The flow has to evolve  for ``infinite time'' to
arrive at the fixed point \cite{wilson1}.
At low energies, the RG flow obtained in the effective theory
should be a good approximation to the
corresponding one of the high-energy theory.
However, within the frame work of the
effective theory, the UV cutoff
$\Lambda_0$ can not become infinite, or
the RG flow does not converge to a fixed point.
Above some scale $\Lambda_{\rm H}$,
the effective theory is no longer effective,
and  should be replaced by
the high-energy theory
to obtain the correct high-energy behavior of the RG flow.

So far there is nothing special. Suppose now
we have 
a trivial theory that well describes
low energy physics,
but we do not know about its high-energy theory.
Our basic assumption is that
the RG flow in the effective theory
that  evolves for ``maximal time''
may be  the best approximation to the renormalized trajectory
of the unknown high-energy theory.
This optimized RG flow can  be obtained
by fine tuning the IR 
values of the dependent parameters of the effective theory.
If the theory is perturbatively 
renormalizable,  we 
regard  the coupling constants with
a negative canonical dimension  as dependent 
parameters. In the case of perturbatively nonrenormalizable theory, 
we  regard  the coupling constants with 
a canonical dimension $d \geq d_{\rm  max} < 0$ as independent parameters, 
while we regard the coupling constants 
with $d < d_{\rm max}$ as dependent 
parameters. (The value of the maximal canonical dimension 
$d_{\rm  max}$ depends on the theory,
and we do not know it a priori.)
Then we require that,
for given low-energy values
of the independent parameters, the low-energy values of the dependent 
parameters are so fine tuned as
to reach the maximal UV cutoff $\Lambda_{\rm max}$. 
The effective theory so optimized
will behave as a local field theory to the shortest distance ($\sim 
\Lambda^{-1}_{\rm max}$).
This is why we would like to call this optimizing
method maximal locality method.
In \cite{Kubo:2001tr} we  considered the uncompactified scalar theories
in higher dimensions, and found
that the maximization 
of the UV cutoff in 
an nonrenormalizable field theory
can give the theory more predictive power,
at least in lower orders of  the local potential approximation
to the exact RG equation.

\subsection{Continuous Wilsonian Renormalization Group}

As we have mentioned above, our interest is
directed to trivial theories. To define such theories in a
non-perturbative fashion, we have to introduce a cutoff. A natural
framework to study cutoff theories is provided by the continuous
Wilsonian RG \cite{wilson1}. Let us briefly illustrate
the basic idea of the Wilsonian RG approach in the
case of $N$ components scalar theory in flat Euclidean $D$
dimensions. As first, we divides the field  $\phi(p)$ in the momentum
space into lower and higher energy modes than the cutoff $\Lambda$
according to
\begin{eqnarray}
\phi^a (p)=\theta(|p|-\Lambda)~\phi^a_> (p) + \theta (\Lambda-|p|) 
\phi^a_< (p) , ~~  a=1,\cdots,N .
\end{eqnarray}
Then the Wilsonian effective action at $\Lambda$ is defined by
integrating out only the higher energy modes $\phi^a_>$ in the path
integral,
\begin{eqnarray}
S_{\rm eff} [\phi_{<},\Lambda] = -\ln \left\{
\int {\cal D}\phi_{>} e^{- S[\phi_{>},\phi_{<}]}
\right\}.
\end{eqnarray}
It was shown that the path integral corresponding to the difference
between $\Lambda$ and $\Lambda + \delta \Lambda$
\begin{eqnarray}
\delta S_{\rm eff} =
S_{\rm eff} [~\phi_{<},\Lambda+\delta\Lambda~] - S_{\rm eff} 
[~\phi_{<},\Lambda~]
\end{eqnarray}
for an infinitesimal $\delta\Lambda$ can be exactly carried out. This
yields the non-perturbative (exact) renormalization group evolution
equation for the effective action
\begin{eqnarray}
\frac{\partial S_{\rm eff}}{\partial t} 
= - \Lambda \frac{\partial S_{\rm eff}}{\partial \Lambda} 
= {\cal O}(S_{\rm eff}) ,
\end{eqnarray}
where ${\cal O}$ is a non-linear operator acting on the functional 
$S_{\rm eff}$. 
There exist various (equivalent) formulations of regularizations, 
but in this paper we consider only the Wegner-Houghton equation
\cite{wegner}. Since $S_{\rm eff}$ is a functional of fields, 
one can think of the  Wegner-Houghton equation as coupled 
differential 
equations for  infinitely many couplings in the effective action. 
The crucial point is that ${\cal O}$
can be exactly derived for a given theory, in contrast to the
perturbative renormalization group approach where the RG equations 
are known only up to a certain order in perturbation
theory. This provides us with possibilities to use approximation 
methods
that go beyond the conventional perturbation theory. Therefore 
Wilsonian
RG approach is suitable for maximal locality
method which deals with non-perturbative effects of 
nonrenormalizable theory.

\subsection{Example: Uncompactified Four-Dimensional Scalar Theory}
\label{four-d}

In this subsection we would like to illustrate how  maximal locality
method works even in a perturbatively renormalizable, but trivial
theory.
We shall consider an uncompactified four-dimensional
scalar theory with four components. The
theory is perturbatively renormalizable, 
but presumably it is trivial \cite{trivial,luscher1,luscher2}.
Here we assume that it is trivial.
Perturbative series are only asymptotic series, and 
suffer from a non-perturbative ambiguity which 
originates from the
renormalon singularity in the Borel plane \cite{renormalon}.
We will see that the  method can remove this ambiguities.

At first, in the derivative expansion approximation,
\cite{nicol,hasenfratz} \cite{wetterich2}\tocite{aoki1} one assumes 
that the effective action $S_{\rm eff}[\phi,\Lambda ]$ can be written as a
space-time integral of a (quasi) local function of $\phi$, i.e.,
\begin{eqnarray}
S_{\rm eff}[\phi,\Lambda ] &=&
\int d^D x \left( \frac{1}{2}\, \sum_{k,l=1}^N
\partial_{\mu}\phi^k\partial_{\mu}\phi^l Z^{kl}(\phi,\Lambda)
+V(\phi,\Lambda)+\cdots \right) ,
\end{eqnarray}
where $\cdots$ stands for terms with higher order derivatives with
respect to the space-time coordinates, and $N$ is a number of scalar
components. In the lowest order of the
derivative expansion (the local potential approximation), there is 
no wave function renormalization, i.e. 
$Z^{kl}(\phi,\Lambda)=\delta^{kl}$, 
and the RG equation for the effective potential $V$ can be obtained. 
Since it is more convenient to work with the RG equation for
dimensionless quantities, which makes the  scaling properties more
transparent, we rescale the quantities according to
\begin{eqnarray}
p \to \Lambda p ,~~ \phi_a \to \Lambda^{\frac{D}{2}-1} \phi_a ,~~ 
V \to \Lambda^{D} V .
\label{rescale}
\end{eqnarray}
Then the RG equation for $V(\phi,\Lambda)$ is given by 
\cite{hasenfratz}
\begin{eqnarray}
\frac{\partial V}{\partial t} 
&=& - \Lambda \frac{\partial V}{\partial \Lambda} \nonumber \\
&=& D V + (2 - D) \rho V 
+ \frac{A_D}{2} \Bigl[ (N-1) \ln (1+V') + \ln (1+V'+2\rho V'') \Bigr] 
\label{WH-V}
\end{eqnarray}
where the prime on $V$ stands for the derivative with respect to 
$\rho$, and
\begin{eqnarray}
\rho = \frac{1}{2} \sum_{a=1}^N \phi_a \phi_a ~,~~ 
A_D = \frac{1}{2^{D-1} \pi^{D/2} \Gamma( D/2 )} .
\label{A_D}
\end{eqnarray}
In the case of $D=4$, we have $A_4 = 1/8\pi^2$. Eq.~(\ref{WH-V}) is the 
Wegner-Houghton equation for the effective 
potential
of the $D$-dimensional $O(N)$ scalar theory. The equation (\ref{WH-V})  is
equivalent to the following equation for $F \equiv \partial V/\partial \rho$,
\begin{eqnarray}
\frac{\partial F}{\partial t} 
= 2 F + (2-D) \rho F' + \frac{A_D}{2} \left[ (N-1) \frac{F'}{1+F} 
+ \frac{3 F' + 2 \rho F''}{1 + F + 2 \rho F'} \right] .
\label{WH-F}
\end{eqnarray}
To solve  eq.~(\ref{WH-F}) in the local potential approximation,
we expand the effective potential $V$ as
\begin{eqnarray}
V(\rho,t) &=& v_0(t) + \sum_{n=1}^{\infty} \frac{1}{n+1}
\frac{f_n(t)}{{(2 A_D)}^n} \Bigl[ \rho - 2 A_D f_0(t) \Bigr]^{n+1} 
\nonumber \\
&=& v_0(t) + \frac{1}{2} \frac{f_1(t)}{2 A_D} {\Bigl[ \rho - 2 A_D 
f_0(0) \Bigr]}^2 
+ \frac{1}{3} \frac{f_2(t)}{(2 A_D)^2} {\Bigl[ \rho - 2 A_D f_0(0) 
\Bigr]}^3 + \cdots .
\label{V-ansatz}
\end{eqnarray}
$F$ can also be expanded, and by inserting expanded 
form of $F$ into   eq.~(\ref{WH-F}),
we can obtain  a set of $\beta$ functions\footnote{The
explicit expressions in lower orders are given in \cite{Kubo:2001tr}.}, 
$\beta_n = d f_n/d t$, at any finite order of
truncation.

Next to see the relation between perturbative
renormalizability and the nonperturbative ambiguity,
we solve the reduction equation,
\cite{reduction}\footnote{Reduction of coupling constants 
has been applied to quantum gravity \cite{atance1} 
and to chiral Lagrangian.\cite{atance2}}
\begin{eqnarray}
\beta_1 \frac{d f_n}{d f_1} & =& \beta_n~,~(n\neq 1)~,
\end{eqnarray}
near the Gaussian fixed point $(f_0=3/4~,~f_n=0 ~(n \geq 1))$
for $D=4$ and $N=4$. 
We find that the general solution for $(n \leq 2)$,
for instance, takes the form
\begin{eqnarray}
f_0 &=&
\frac{3}{4}- \frac{9}{16}f_1^2 + \frac{225}{64} f_1^2 - 
\frac{7857}{256}f_1^3 + 
\frac{269001}{1024 }f_1^4- \frac{12806991}{4096}f_1^5 + 
 \frac{650870883}{16384}f_1^6+O(f_1^7) \nonumber \\
 & &
+ K_2\exp \left( -\frac{2}{3f_1}- \frac{57}{4}f_1 \right)~
f_1^{7/2}~\left[ \frac{3}{16} - \frac{9}{8 } f_1 
+ \frac{2799}{256}f_1^2+O(f_1^3)~\right]~,
\label{general0} \\
f_2 &=& 
\frac{15}{4} f_1^3 - \frac{189}{8 }f_1^4+ 
\frac{7479}{64 }f_1^5- \frac{12879}{32 }f_1^6+O(f_1^7)
+ K_2 \exp \left( -\frac{2}{3f_1}- \frac{57}{4}f_1 \right)~
f_1^{9/2}~, \nonumber \\
\label{generaln}
\end{eqnarray}
where $K_2$ is an integration constant.
We see from the above solution that  the exponential function 
$\exp{(-2/3f_1)}$ decreases fast
as $f_1$ approaches zero, so that the ambiguity
involved in the integration constant $K_2$ becomes negligible
in the infrared limit.
In this limit, 
the power series part of 
the above solution (\ref{generaln}) becomes dominant.
In other words, the power series solution
is infrared attractive.
This infrared attractiveness is 
interpreted as perturbative 
renormalizability by Polchinski \cite{polchinski}. 

As we will argue below, 
the exponential part of (\ref{generaln})
is a non-perturbative ambiguity \cite{luscher1}.
We have computed higher orders in
the power series expansion (\ref{generaln})  and found that they 
do not approximate the exact (numerical) result better. 
The one with the first four terms in (\ref{generaln})  
is the best approximation among lower orders.
From this fact, we believe that 
the power series solution (\ref{generaln}) does not converge,
and that it is an asymptotic series.
So, this power series reflects the property of 
perturbation series in the conventional perturbation theory, 
as far as our numerical analysis in lower orders suggests.
This interpretation is also supported by the fact that not only 
the leading form of the nonperturbative ambiguity, 
the last exponential term in (\ref{generaln}), agrees with 
that of the known renormalon ambiguity \cite{renormalon}, 
but also the coefficient of $1/f_1$,  $2/3$, in the exponential function.
The power of $f_1$ in front of the exponential function, 
that is $9/2$, differs slightly from the expected value $3$.
The origin is presumably the local potential approximation 
to the exact RG evolution equation. We believe that 
the last term in (\ref{generaln}) is the renormalon ambiguity.

According to the formulation of our method in Sect.~\ref{formulation}, 
we should regard $f_0$ and $f_1$ as independent parameters, while 
the other coupling constants $f_n (n \ge 2)$ as dependent parameters. 
Then, using a set of $\beta$ functions for each coupling constants, 
we investigate the running time $T_0 = \ln(\Lambda_0/\Lambda)$ 
against $f_n(t=0) \ (n \ge 2)$ for given $f_0(0)$ and $f_1(0)$. 
\begin{figure}
 \centerline{\includegraphics[width=0.5\linewidth]
   {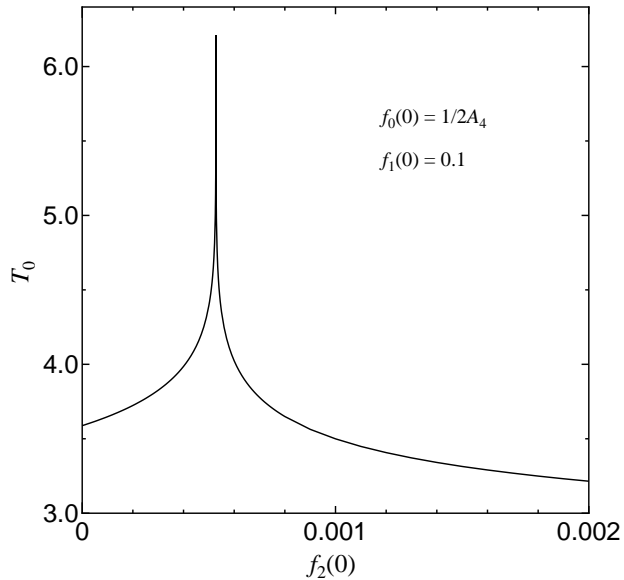}}
  \caption{Fine-tuning of $f_2(0)$ for given values 
     $(f_0(0),~f_1(0)) = (1/2A_4,~0.10)$ 
     in uncompactified four-dimensional case for the 
     truncation at $n=2$. The running time $T_0 = 
     \ln(\Lambda_0/\Lambda)$ is peaked at $f_2(0) \simeq 0.000528$.}
  \label{4D-peak1}
\end{figure}
\begin{figure}
 \centerline{\includegraphics[width=0.7\linewidth]
   {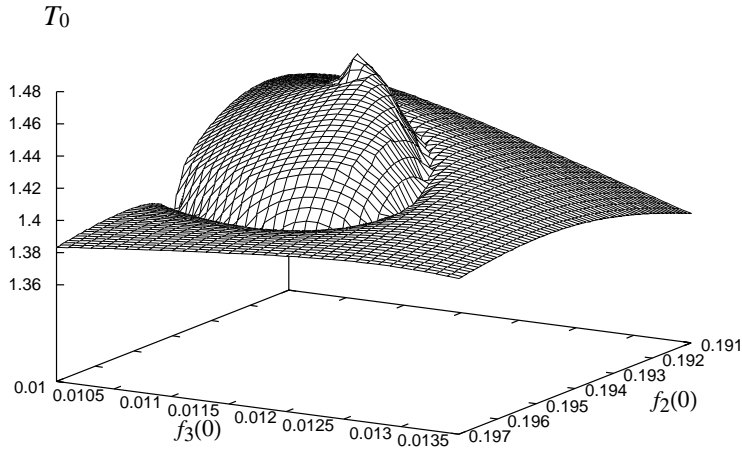}}
  \caption{Fine-tuning of $f_2(0)$ and $f_3(0)$ for
     $(f_0(0),~f_1(0)) = (1/2A_4,~0.79)$ 
     in the case of truncation at $n=3$. 
     $T_0$ is peaked at $(f_2(0),~f_3(0)) \simeq (0.19285,~0.0112)$ .}
  \label{4D-peak2}
\end{figure} \\
Fig.~\ref{4D-peak1} and Fig.~\ref{4D-peak2} show the results for given 
values,
$(f_0(0),~f_1(0)) = (1/2A_4,~0.10)$, in the case of the truncation 
level at $n=2$, and for given $(f_0(0),~f_1(0)) = (1/2A_4,~0.79)$ at $n=3$
respectively. From these figures, we can see that $T_0$ is peaked at
$f_2(0) \simeq 0.000528$ for $n=2$, and peaked at
$(f_2(0),~f_3(0)) \simeq (0.19285,~0.0112)$ for $n=3$. Then 
maximal locality method requires that the dependent coupling constants 
$f_n~(n \ge 2)$ must be so fine tuned that $T_0$ becomes maximal. 
In this way we can determine the values of the dependent parameters 
for a given finite number of the independent parameters. 
\begin{figure}
 \centerline{\includegraphics[width=0.5\linewidth]
   {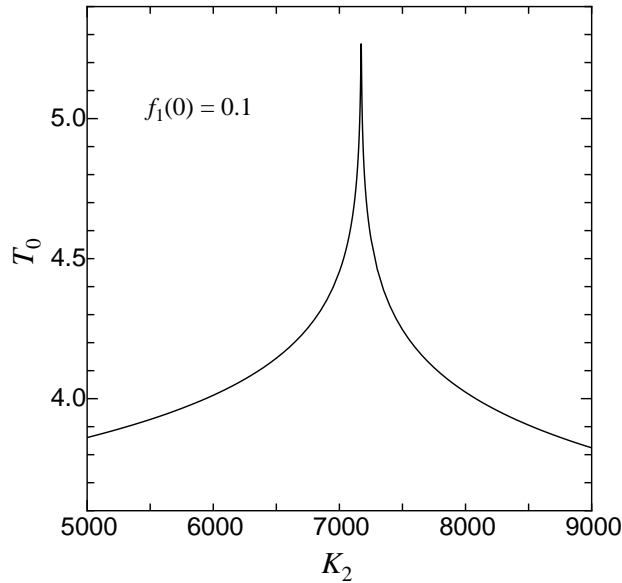}}
  \caption{Determination of the nonperturbative coefficient $K_2$
   (given in (\ref{generaln}))  in $D=4$.}
  \label{K2det}
\end{figure}
This implies that 
the constant $K_2$ in (\ref{generaln}), which exhibits
a nonperturbative correction of the renormalon 
type \cite{renormalon,luscher1}, is also fixed.
In Fig.~\ref{K2det} we plot
$T_0$ against $K_2$. From this result we obtain
\begin{eqnarray}
K_{2} &\simeq  7 \times 10^3~.
\label{k2}
\end{eqnarray}
This means a departure of about  $3~(0.1)$ \%  from the perturbative result
at $f_1=0.1~(0.07)$. Needless to say that  the
corresponding effect  in the standard model could be in principle measurable.


\section{Application to the Compactified Theories}
\label{Appli}

\subsection{The Action}

We now come to consider the Kaluza-Klein theory for the $N$ components
scalar field in Euclidean $D+1$ dimensions where we assume that the 
one
dimension is compactified on a circle with radius $R$. We denote 
the
one compactified coordinate by $y$ and other flat coordinates by $x_M$
$(M = 0,1,\cdots,D)$. 

We start with the following  action in $(D+1)$-dimensions:
\begin{eqnarray}
S_{D+1} 
= \int d^D x \int^{2 \pi R}_0 d y 
\left\{ \frac{1}{2} \sum_a^N \left( {(\partial_M \Phi_a)}^2 
+ {(\partial_y \Phi_a)}^2 \right) + V_{D+1}(\Phi_b) \right\} ,
\end{eqnarray}
where $V_{D+1}$ is the $(D+1)$-dimensional potential.
As eq.~(\ref{V-ansatz}), $V$  is assumed to be
expanded as
\begin{eqnarray}
V_{D+1} 
= v_0(t) + \sum_{m=1}^{\infty} \frac{f_m(t)}{(2 A_{D+1})^m} 
\left[ \frac{1}{2} \sum_a \Phi_a \Phi_a - 2 A_{D+1} f_0(t) 
\right]^{m+1} .
\label{D+1-pot}
\end{eqnarray}
The scalar field $\Phi$ satisfies the boundary condition on the extra
coordinate $y$,
\begin{eqnarray}
\Phi_a(x,y) = \Phi_a(x,y + 2\pi R) \ .
\end{eqnarray}
Then the field can be expanded as
\begin{eqnarray}
\Phi_a(x,y) = \sum_{n \in {\bf Z}} e^{i n \frac{y}{R}} 
\phi^{(n)}_a(x) ~.
\end{eqnarray}
$\phi^{(n)}$ is $n$-th Kaluza-Klein mode. After we appropriately 
perform
rescaling to the field and coupling constants, and integrate out only
$y$ coordinate of extra dimension,  we obtain the $D$-dimensional
potential
\begin{eqnarray}
V_D &=& {v_0}' (t) + \frac{1}{2} \left( f_0^2 f_2 - f_0 f_1 \right) 
\sum_a \sum_n \phi_a^{(n)} \phi_a^{(-n)} \nonumber \\
&& + \frac{1}{8 A_D} \left( \frac{1}{2} f_1 - f_0 f_2 \right) 
\sum_{a,b} \sum_{n_i} 
\phi_a^{(n_1)} \phi_a^{(n_2)} \phi_b^{(n_3)} \phi_b^{(n_4)} 
~\delta_{n_1 + n_2 + n_3 + n_4,~0} \nonumber \\
&& + \frac{1}{24 (2 A_D)^2} f_2 \sum_{a,b,c} \sum_{n_i} 
\phi_a^{(n_1)} \phi_a^{(n_2)} \phi_b^{(n_3)} 
\phi_b^{(n_4)} \phi_c^{(n_5)}  \phi_c^{(n_6)} 
~\delta_{n_1 + n_2 + n_3 + n_4 + n_5 + n_6,~0} \nonumber \\
&& + \cdots ,
\label{KK-pot}
\end{eqnarray}
and the action in  $D$-dimensions
\begin{eqnarray}
S_{D} 
= \int d^D x \left\{ \frac{1}{2} \sum_{a=1}^N \sum_{n \in {\bf Z}} 
\phi^{(-n)}_a \left( - \partial_M^2 + m_n^2 \right) \phi^{(n)}_a 
+ V_D(\phi^{(l)}_b) \right\} ,
\end{eqnarray}
where $m_n$ is $n$-th mode Kaluza-Klein masses,
\begin{eqnarray}
m_n = \frac{n}{R} \ .
\end{eqnarray}
As we have seen, the compactified $(D+1)$-dimensional theory yields
an infinite number of Kaluza-Klein modes at the viewpoint of the flat 
$D$-dimensional theory.

\subsection{Wegner-Houghton Equation for Compactified Scalar Theories}

Now we would like to investigate the Wegner-Houghton equation for the
compactified $(D+1)$-dimensional scalar theory in 
the $D$-dimensional
sense. In the lowest order of derivative expansion approximation,
we assume the following effective action for the compactified theory
\begin{eqnarray}
S_{\rm eff} [\phi^{(l)}_b,t] 
= \int d^D x \left\{ \frac{1}{2} \sum_{a=1}^N \sum_{n \in {\bf Z}} 
\phi^{(-n)}_a \left( - \partial_M^2 + m_n^2 \right) \phi^{(n)}_a 
+ V_{\rm eff}(\phi^{(l)}_b,t) \right\} .
\end{eqnarray}
Then the RG equation for the effective potential $V_{\rm eff}$ can be
obtained. As in Sect.~\ref{four-d},  we rescale the quantities 
according
to eq.~(\ref{rescale}) and
\begin{eqnarray}
m_n \to \Lambda m_n .
\label{rescale2}
\end{eqnarray}
The RG equation (Wegner-Houghton equation) for the effective potential $V$ is
given by\footnote{In the following, we rewrite the effective potential
by $V$.}
\begin{eqnarray}
\frac{\partial V}{\partial t} 
&=& - \Lambda \frac{\partial V}{\partial \Lambda} \nonumber \\
&=& D V - \frac{D-2}{2} \sum_n \sum_a \phi_a^{(n)} 
\frac{\partial V}{\partial \phi_a^{(n)}} 
+ \frac{A_D}{2} {\rm Tr} \ln \left( 1 + m_n^2 
+ \frac{\partial^2 V}{\partial \phi_a^{(n)} \partial \phi_b^{(-n)}} 
\right) ,
\label{KK-WH-V}
\end{eqnarray}
where the trace of the second term in the right hand side 
means summation over
the flavor indices $a$ and the Kaluza-Klein mode indices $n$, 
and $m_n$ is rescaled
dimensionless Kaluza-Klein masses defined in eq.~(\ref{rescale2}). 
Note that we have considered only the diagonal parts of 
the Kaluza-Klein indices in the logarithmic function of 
the right hand side. Off-diagonal parts in the logarithmic
function yield vertices that depend on the external Kaluza-Klein 
modes, and so this is beyond the local potential approximation, 
since Kaluza-Klein indices can be regarded as fifth momentum of the field.

Eq.~(\ref{KK-WH-V}) is the central equation that we will analyze in the
following. Therefore, all the results we will obtain are
valid only within the local potential approximation. From the
next section, we would like to investigate the predicted values of
the dependent parameters in  compactified five-, six-, seven- and 
eight-dimensional scalar theories, and to check the consistency of the
predictions from these theories and the uncompactified flat theories.

\section{Results}
\label{results}

\subsection{Compactified Five-Dimensional Case}

We first consider a compactified five-dimensional scalar 
theory with four components, 
where one extra dimension is compactified on a circle with radius $R$
and other four dimensions are uncompactified. 
We would like to find out whether the
uncompactified four-dimensional scalar theory 
can be regarded as the effective theory of the
compactified five-dimensional theory, 
if we apply maximal locality method to the compactified
five-dimensional theory as well as to the uncompactified
four-dimensional theory.
To this end, 
we 
make predictions on the dependent
parameters at low energy 
scale
$\Lambda_{\rm R} (\ll R^{-1})$ by 
applying maximal locality method to
the compactified theory.
We then compare them with those
obtained in the uncompactified four-dimensional theory.
If it is the low-energy effective theory of the compactified five-dimensional 
theory, the predicted values should agree with each other.

As in eq.~(\ref{D+1-pot}), we start with 
$4+1(=5)$-dimensional
potential,
\begin{eqnarray}
V_{4+1(=5)} 
= v_0(t) + \sum_{m=1}^{\infty} \frac{f_m(t)}{(2 A_5)^m} 
\left[ \frac{1}{2} \sum_a \Phi_a \Phi_a - 2 A_5 f_0(t) \right]^{m+1} ,
\label{4+1-pot}
\end{eqnarray}
which defines the coupling constants $f_m$. 
After integrating  out only the $y$ coordinate,
we appropriately perform rescaling to the field and coupling
constants, then we obtain the potential in terms of four-dimensional 
theory like eq.~(\ref{KK-pot}).
Then we can obtain a set of $\beta$-functions  $\beta_m = df_m/dt$ 
from
eq.~(\ref{KK-WH-V}) at any finite order truncation. The explicit
expressions in lower orders are given in Appendix
\ref{KK-beta}.\footnote{These $\beta$-functions can be obtained by
comparing at each order of ${(\frac{1}{2}\sum_a \phi_a^{(0)}
\phi_a^{(0)} - 2 A_D f_0)}^n$ in eq.~(\ref{KK-WH-V}), 
since we are interested in the behavior of the
coupling constants of the Kaluza-Klein zero-modes.} 
From these explicit expressions, we can see that these
$\beta$-functions approach the flat four-dimensional forms 
as $R \to 0$.
\begin{figure}
 \centerline{\includegraphics[width=0.5\linewidth]
   {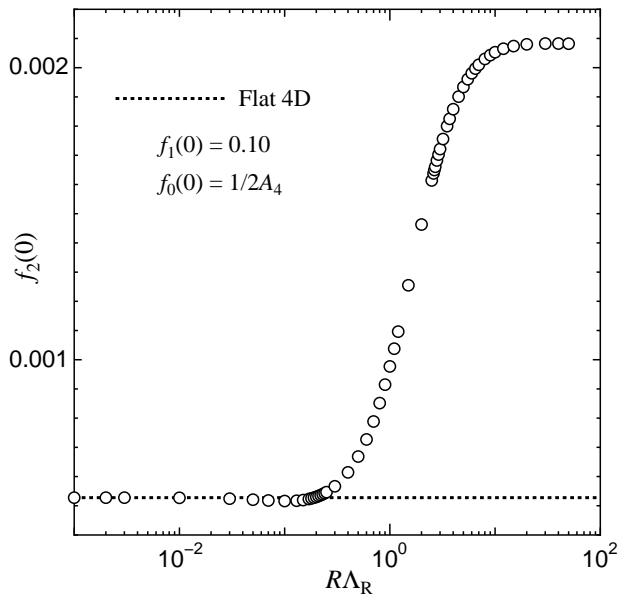}}
  \caption{Predicted values of $f_2(0)$ for 
   $(f_0(0),~f_1(0)) = (1/2A_4,~0.10)$ in compactified five-dimensional 
   theory. The dotted line shows the value in flat four-dimensional 
   theory.}
  \label{pr-com5D}
\end{figure}

As in the flat four-dimensional case in
Sect.~\ref{formulation}, we have to regard the coupling constants 
$f_0$ and $f_1$ as independent parameters, 
and other coupling constants $f_m~(m \ge 2)$ as dependent parameters. 
For the sake of simplicity, we calculate the predicted
values of $f_2(0)$ at truncation order $m=2$. 
Fig.~\ref{pr-com5D} shows the predicted values of $f_2(0)$ for
$(f_0(0),~f_1(0)) = (1/2A_4,~0.10)$ 
as a function of compactification scale 
$R^{-1}$ varying
from $10^{-2} \times \Lambda_{\rm R}$ to $10^3 \times \Lambda_{\rm 
R}$, where $\Lambda_{\rm R}$ is 
renormalization scale (i.e. 
$t=0$ corresponds to $\Lambda=\Lambda_{\rm R}$).  We can see from this
figure that the predicted values almost do not differ  up to 
$R\Lambda_{\rm R}
\sim 0.1$ from the four-dimensional ones. This means that at
most up to $R\Lambda_{\rm R} \sim 0.1$, namely $R^{-1} \gsim 10 \times
\Lambda_{\rm R}$, the consistency is ensured.
This result can also give us the bound for the compactification scale
$R^{-1}$.
Since  the renormalization scale $\Lambda_{\rm R}$ can
represent a typical energy scale, it is natural to identify
with the Higgs's VEV.
Then we  find 
\begin{eqnarray}
R^{-1} \gsim {\rm {\cal O} (TeV)}.
\end{eqnarray}
Of course, this bound can change if the value
of the independent parameter 
$f_1(0)$ changes. The change will be investigated  in
Sect.~\ref{diverse-f1}.

\subsection{Compactified Six-, Seven- and Eight-Dimensional Case}

Here we consider compactified six-, seven- and eight-dimensional 
scalar
theories. 
The situation here is different from the previous five-dimensional case, 
because in the previous case the effective theory was perturbatively
renormalizable. 
In the cases at hand, 
the compactified as well as lower-dimensional 
uncompactified theories are nonrenormalizable.
We investigate the prediction of the dependent
coupling $f_2$ at truncation order $m=2$ for various compactification 
scales.  
\begin{figure}
 \centerline{\includegraphics[width=0.5\linewidth]
   {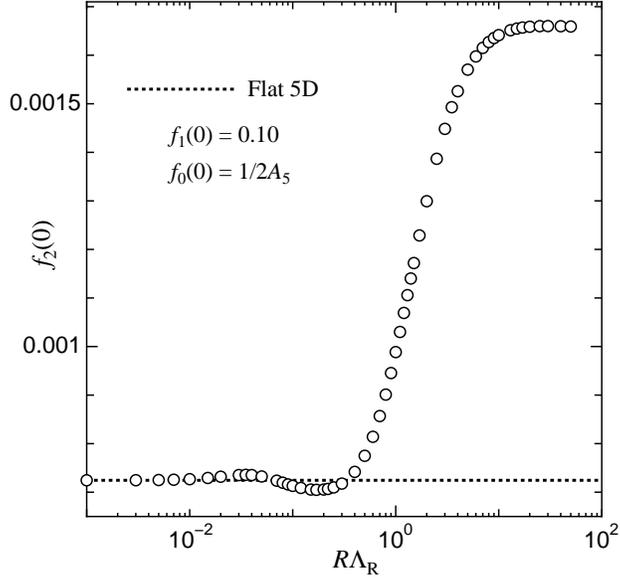}}
  \caption{The same as Fig.~\ref{pr-com5D} for 
   $(f_0(0),~f_1(0)) = (1/2A_5,~0.10)$ 
   in compactified six-dimensional theory.}
  \label{pr-com6D}
\end{figure}
\begin{figure}
 \centerline{\includegraphics[width=0.5\linewidth]
   {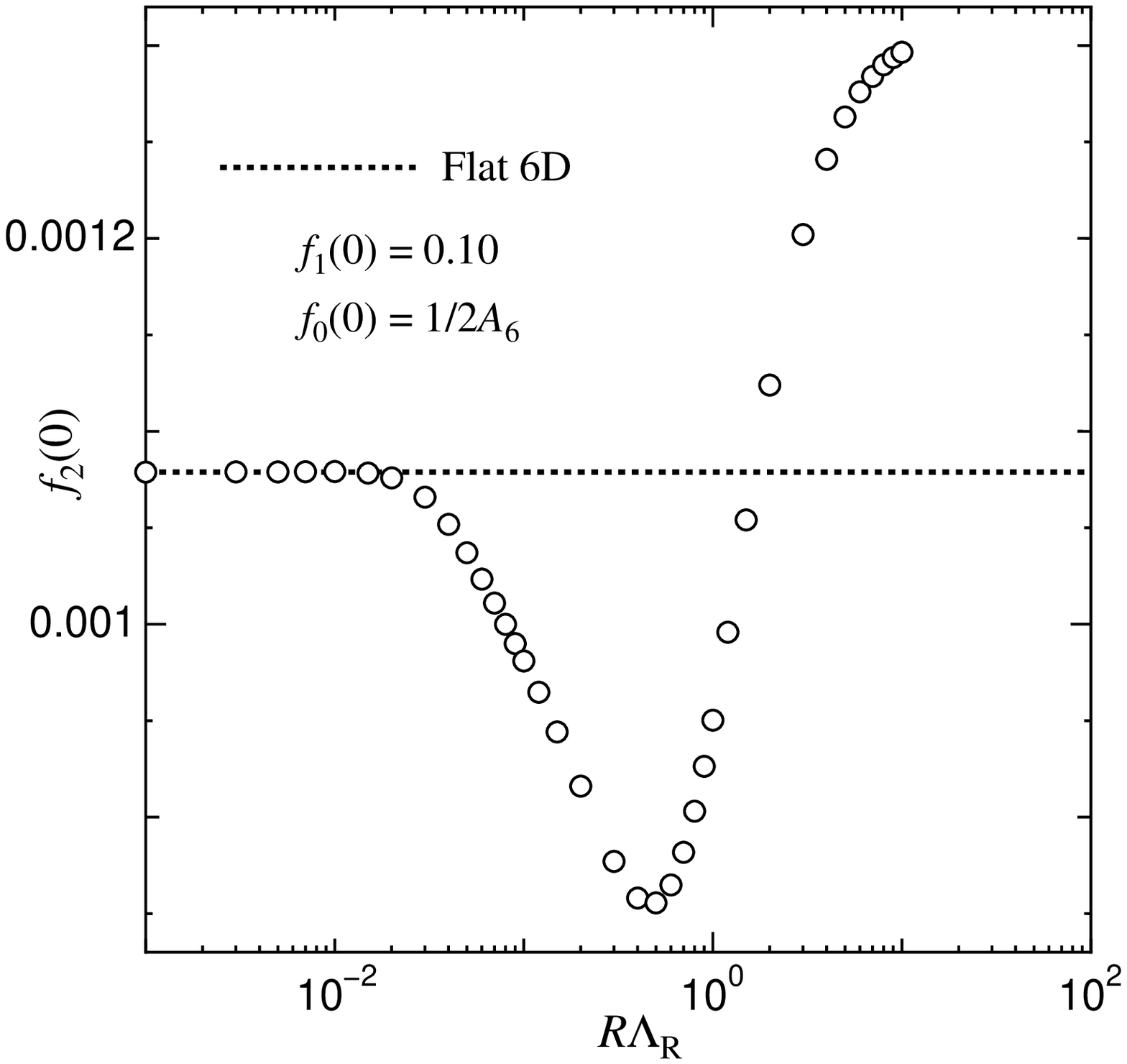}}
  \caption{The same as Fig.~\ref{pr-com5D} for 
   $(f_0(0),~f_1(0)) = (1/2A_6,~0.10)$ 
   in compactified seven-dimensional theory.}
  \label{pr-com7D}
\end{figure}
\begin{figure}
 \centerline{\includegraphics[width=0.5\linewidth]
   {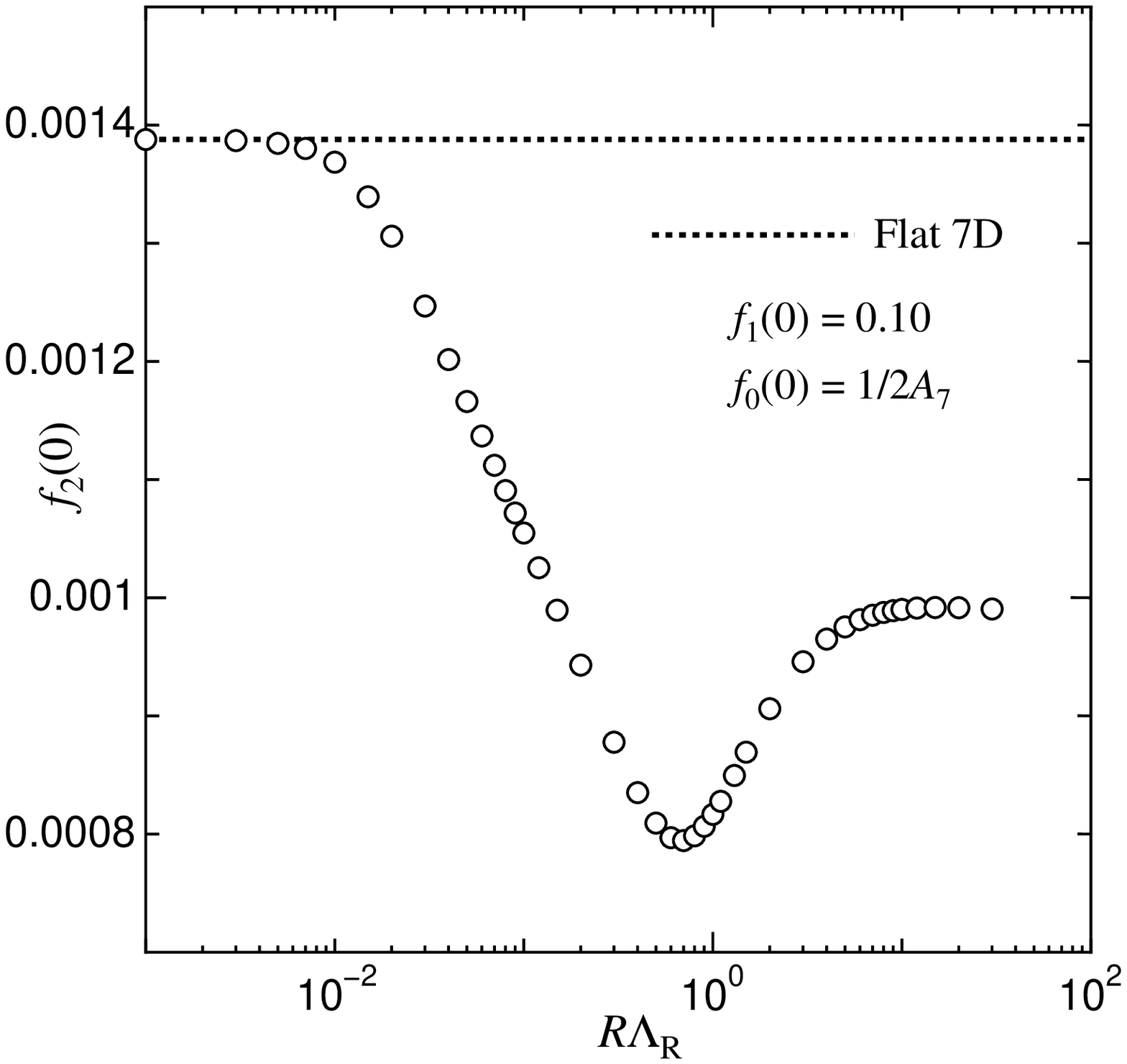}}
  \caption{The same as Fig.~\ref{pr-com5D} for 
   $(f_0(0),~f_1(0)) = (1/2A_7,~0.10)$ 
   in compactified eight-dimensional theory.}
  \label{pr-com8D}
\end{figure}
The results are shown 
in Fig.~\ref{pr-com6D}, \ref{pr-com7D} and
\ref{pr-com8D}. To calculate the value $f_2(0)$, we have used
$(f_0(0),~f_1(0)) = (1/2A_D,~0.10)$.
From first two  figures, 
we found that 
\begin{eqnarray}
R^{-1} \gsim 50 \times \Lambda_{\rm R} ,
\end{eqnarray}
is the consistency bound
in six and seven dimension cases.
This means that if the consistency bound is satisfied,
we
can predict at low-energies the dependent parameters 
of the compactified original theory 
within the framework of the lower-dimensional uncompactified
theory by using maximal locality method.

Finally, we show the  eight-dimensional case in
Fig.~\ref{pr-com8D}. We can see from the figure that if
compactification scale $R^{-1}$  satisfies the condition
\begin{eqnarray}
R^{-1} \gsim 100 \times \Lambda_{\rm R} ,
\label{7result}
\end{eqnarray}
the predictions of the compactified and uncompactified theories
agree with each other.

\subsection{Six-Dimensional Case in Diverse Independent Parameter 
$f_1$}
\label{diverse-f1}

Until to now, we have assume the  same
value of the independent coupling constant $f_1$ at renormalization 
scale
$\Lambda_{\rm R}$, i.e.,  $f_1(0)=0.10$. The prediction
of the dependent coupling constants change if the
value of $f_1(0)$ changes. 
Here we would like to calculate  the change
as a function of the independent coupling constant
$f_1$. 
\begin{figure}
 \centerline{\includegraphics[width=0.9\linewidth]
   {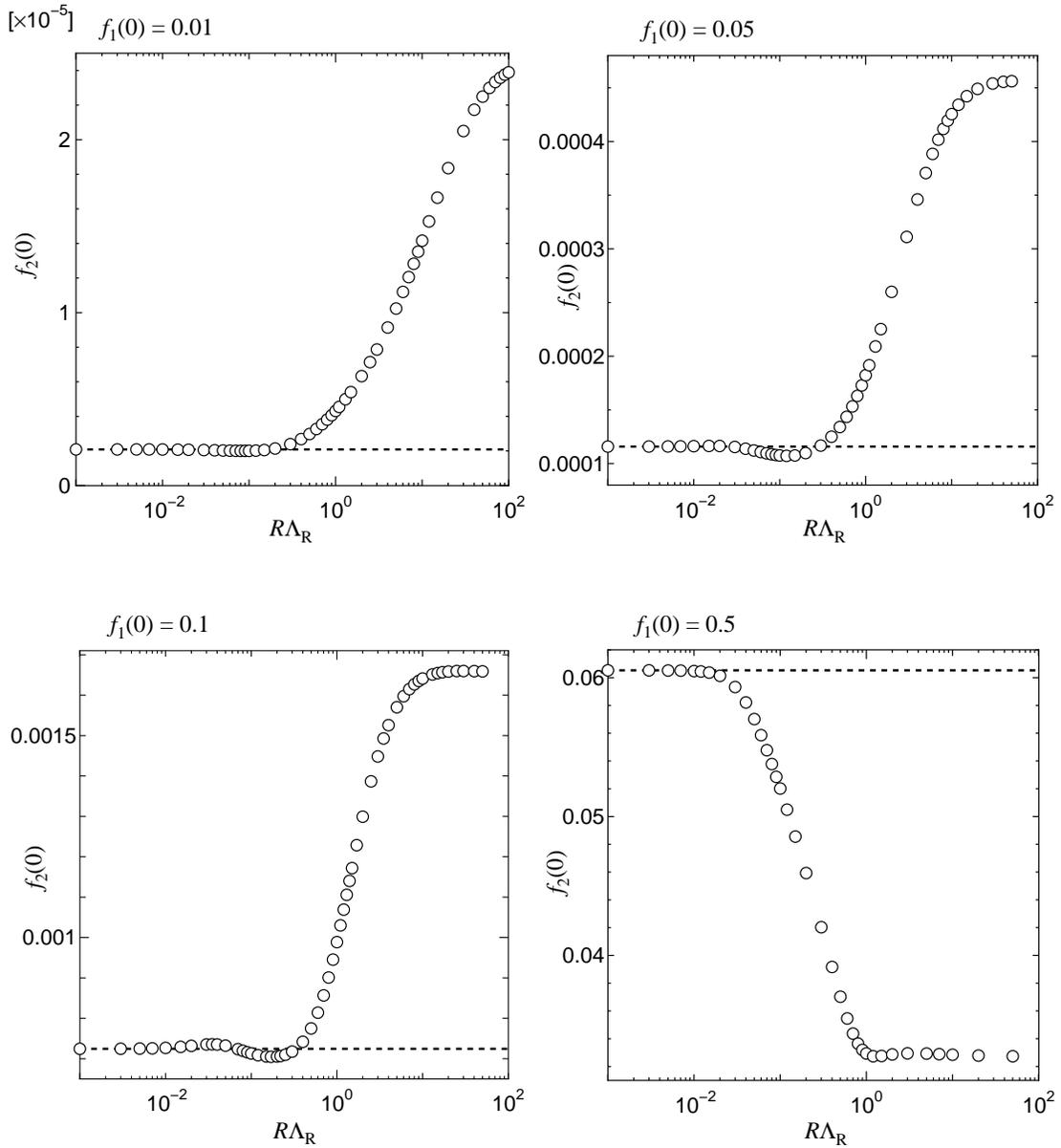}}
  \caption{The predictions of $f_2(0)$ in the compactified six-dimensional 
   theory for diverse independent parameter $f_1(0)$. All figures plot for 
   $f_0(0)=1/2A_5$, and the dotted line shows the predicted values in the 
   uncompactified five-dimensional theory.}
  \label{6D-divf1}
\end{figure}
To this end, we consider the compactified six-dimensional 
scalar theory of the previous subsection, and calculate 
the consistency bound for the compactification scale 
as a function of the  independent parameter
$f_1(0)$. In Fig.~\ref{6D-divf1}, we show the results for 
$f_1(0) = 0.01,~0.05,~0.10$
and $0.50$ for the same  renormalization condition $f_0(0) = 1/2 A_5$. 
As we can see from these figures, 
the point of $R \Lambda_{\rm R}$ at which the predicted values 
start to separate from each other,  becomes smaller  as
$f_1(0)$ increases. 
We, therefore, may conclude that 
as far as $f_1(0) \lsim 0.5$,
the compactification scale $R^{-1}$ has to satisfy
\begin{eqnarray}
R^{-1} \gsim 50 \times \Lambda_{\rm R}
\label{bound-R-5}
\end{eqnarray}
for maximal locality method to consistently work.

\section{Conclusion}
\label{conclusion}

In particle physics, perturbatively nonrenormalizable theories 
have played
an important role. They are regarded as low-energy effective theories of
more fundamental, high-energy theories. 
Quantum corrections in a nonrenormalizable theory explicitly depends
on the UV cutoff, and 
infinitely many independent parameters can be generated.
Nonrenormalizable theories have much less
predictive power compared with renormalizable theories, as well known.
Field theories  in
more than four space-time dimensions are usually nonrenormalizable, too.
In recent works on extra
dimensions, the length scale of extra dimensions is
often assumed
to be so large that not only the existence of 
extra dimensions but also
quantum corrections could be experimentally observed. 

In our previous work \cite{Kubo:2001tr}, we
applied the Wilsonian RG to nonrenormalizable
theories, and proposed a method to 
give more predictive power to these theories. 
The
method is
based on the assumption on the existence of maximal 
UV cutoff in a nonrenormalizable
theory, and on the requirement that the dependent, low-energy 
parameters of the
theory should be so
adjusted that one arrives at a maximal cutoff.
A nonrenormalizable cutoff theory, so optimized,   behaves
 as  a local field theory as much as possible.
In the present work, we considered 
$(D+1)$-dimensional
scalar theories with one extra dimension  compactified on a 
circle. 
It is naively expected that 
the uncompactified, flat $D$-dimensional theory is the low-energy 
effective
theory of the compactified $(D+1)$-dimensional theory. 
We asked ourself, whether or not
this expectation is correct,
when maximal locality method is employed  both in
the uncompactified $D$-dimensional and 
compactified $(D+1)$-dimensional theories.
We investigated this question using compactified 
five-,
six-, seven- and eight-dimensional  scalar theories with
four components.
The main finding is
that this consistency requirement can strongly constrain the compactification 
scale $R^{-1}$. 
We found that 
for the consistency requirement to be satisfied,
the compactification scale $R^{-1}$ should be
larger than $\Lambda_{\rm min}=R^{-1}_{\rm max}$, which, depending on 
the dimension $D$, is
$10$ to $100$  times as high as 
the renormalization scale $\Lambda_{\rm R}$ of the effective theory, a typical 
energy scale of the low-energy theory.
Although this
condition for $R$ has been obtained
in the derivative expansion approximation in the 
lowest order to the Wegner-Houghton equation
(\ref{KK-WH-V}), 
we believe that the gross feature does not depend on the approximation 
and regularization schemes used.

Finally, we would like to comment on the relation between our method 
and renormalizability of perturbatively nonrenormalizable theories such as
quantum gravity and higher-dimensional Yang-Mills theory.
The existence of an UV-fixed point means renormalizability of the theory 
according to Weinberg.\cite{weinberg} In Ref.~\citen{QG1}, the exact RG
equation approaches have been applied to Einstein's theory of gravity. 
It has been claimed that within the approximation
used in Ref.~\citen{QG2} there seems to exist an UV-fixed point in the
theory.
Furthermore, the existence of a continuum limit and an UV-fixed point in
Yang-Mills theories in more than four dimensions have been investigated, 
by lattice Monte-Carlo simulations \cite{Ejiri:2000fc} 
and by Wilsonian RG approaches \cite{Gies:2003ic}. 
Those results indicate that, even if the UV fixed point does not exist, 
Einstein's theory and compactified higher-dimensional Yang-Mills theories 
can behave as a local field theory to very short distances.
Therefore, these theories may have
a built-in mechanism to maximize the UV cutoff.
We would like to leave the study on
this issue to future work.

\section*{Acknowledgements}

We would like to K-I.~Aoki and H.~Terao for useful discussions.

\appendix
\section{$\beta$-functions for the Compactified Theory}
\label{KK-beta}

We give here the $\beta$-functions of the coupling constants $f_m$ ($m
\le 3$) for $(D+1)$-dimensional four components scalar theory. 
These functions are described in terms of $D$-dimensional theory, 
and $R$ is compactification radius.

{\small 
\begin{eqnarray*}
\beta_0 
&=& (D-2) f_0 - \frac{3}{4} \pi R \Lambda \coth(\pi R \Lambda) \\
&& - \left( \frac{3}{4} + \frac{f_0 f_2}{f_1} \right) 
\pi R \Lambda \coth{\left( \pi R \Lambda \sqrt{1 + 2 f_0 f_1} 
\right)} 
\frac{1}{\sqrt{1 + 2 f_0 f_1}} \\
\ \\
\beta_1 
&=& (4-D) f_1 - \frac{3}{4} f_1^2 
\left[ \frac{1}{2} \pi R \Lambda \left( \coth(\pi R \Lambda) 
+ \pi R \Lambda {\rm \,csch}^2(\pi R \Lambda) \right) \right] \\
&& + \left( f_2 - 2 \frac{f_0 f_2^2}{f_1} + 3 f_0 f_3 \right) 
\pi R \Lambda \coth{\left( \pi R \Lambda \sqrt{1 + 2 f_0 f_1} 
\right)} 
\frac{1}{\sqrt{1 + 2 f_0 f_1}} \\
&& - \left( \frac{9}{4} f_1^2 + 6 f_0 f_1 f_2 + 4 f_0^2 f_2^2 \right) 
\\
&& \times \left[ \frac{1}{2} \pi R \Lambda 
\left( \frac{1}{\sqrt{1 + 2 f_0 f_1}} 
\coth{\left( \pi R \Lambda \sqrt{1 + 2 f_0 f_1} \right)} 
+ \pi R \Lambda {\rm \,csch}^2 
\left( \pi R \Lambda \sqrt{1 + 2 f_0 f_1} \right) \right) \right] 
\frac{1}{1 + 2 f_0 f_1} \\
\ \\
\beta_2 
&=& (6-2 D) f_2 - \frac{9}{4} f_1 f_2 
\left[ \frac{1}{2} \pi R \Lambda \left( \coth(\pi R \Lambda) 
+ \pi R \Lambda {\rm \,csch}^2(\pi R \Lambda) \right) \right] \\
&& + ~\frac{3}{4} f_1^3 \left[ \frac{3}{8} \pi R \Lambda 
\left( \coth(\pi R \Lambda) 
+ \pi R \Lambda {\rm \,csch}^2(\pi R \Lambda) \right) 
+ \frac{1}{4} {(\pi R \Lambda)}^3 \coth(\pi R \Lambda) 
{\rm \,csch}^2(\pi R \Lambda)  \right] \\
&& + \left( 3 f_3 - \frac{3 f_0 f_2 f_3}{f_1} + 6 f_0 f_4 \right) 
\pi R \Lambda \coth{\left( \pi R \Lambda \sqrt{1 + 2 f_0 f_1} 
\right)} 
\frac{1}{\sqrt{1 + 2 f_0 f_1}} \\
&& - \left( \frac{45}{4} f_1 f_2 + 15 f_0 f_2^2 
+ \frac{27}{2} f_0 f_1 f_3 + 18 f_0^2 f_2 f_3 \right) \\
&& \times \left[ \frac{1}{2} \pi R \Lambda 
\left( \frac{1}{\sqrt{1 + 2 f_0 f_1}} 
\coth{\left( \pi R \Lambda \sqrt{1 + 2 f_0 f_1} \right)} 
+ \pi R \Lambda {\rm \,csch}^2\left( \pi R \Lambda 
\sqrt{1 + 2 f_0 f_1} \right) \right) \right] \frac{1}{1 + 2 f_0 f_1} 
\\
&& + \left( \frac{27}{4} f_1^3 + 27 f_0 f_1^2 f_2^2 
+ 36 f_0^2 f_1 f_2^2 + 16 f_0^3 f_2^3 \right) \\
&& \times \left[ \frac{3}{8} \pi R \Lambda 
\left( \frac{1}{\sqrt{1 + 2 f_0 f_1}} 
\coth{\left( \pi R \Lambda \sqrt{1 + 2 f_0 f_1} \right)} 
+ \pi R \Lambda {\rm \,csch}^2\left( \pi R \Lambda 
\sqrt{1 + 2 f_0 f_1} \right) \right) \frac{1}{{(1 + 2 f_0 f_1)}^2} 
\right. \\
&& ~~~~ \left. + \frac{1}{4} {(\pi R \Lambda)}^3 \coth 
\left( \pi R \Lambda \sqrt{1 + 2 f_0 f_1} \right) 
{\rm \,csch}^2\left( \pi R \Lambda \sqrt{1 + 2 f_0 f_1} \right) 
\frac{1}{{(1 + 2 f_0 f_1)}^{\frac{3}{2}}} \right] \\
\ \\
\beta_3 
&=& (8-3D) f_3 - \left( 3 f_1 f_3 + \frac{3}{2} f_2^2 \right) 
\left[ \frac{1}{2} \pi R \Lambda \left( \coth(\pi R \Lambda) 
+ \pi R \Lambda {\rm \,csch}^2(\pi R \Lambda) \right) \right] \\
&& - ~ \frac{3}{4} f_1^4 \left[ \frac{5}{16} \pi R \Lambda 
\left( \coth(\pi R \Lambda) + \pi R \Lambda 
{\rm \,csch}^2(\pi R \Lambda) \right) 
+ \frac{1}{4} {(\pi R \Lambda)}^3 \coth(\pi R \Lambda) 
{\rm \,csch}^2(\pi R \Lambda) \right. \\
&& ~~~~~~~~~ \left. + \, {(\pi R \Lambda)}^4 
\left( \frac{1}{12} \coth^2(\pi R \Lambda) {\rm \,csch}^2(\pi R 
\Lambda) 
+ \frac{1}{24} {\rm \,csch}^4(\pi R \Lambda) \right) \right] \\
&& + \left( 6 f_4 - \frac{4 f_0 f_2 f_4}{f_1} 
+ 10 f_0 f_5 \right) \pi R \Lambda 
\coth{\left( \pi R \Lambda \sqrt{1 + 2 f_0 f_1} \right)} 
\frac{1}{\sqrt{1 + 2 f_0 f_1}} \\
&& - \left( \frac{25}{2} f_2^2 + 21 f_1 f_3 + 58 f_0 f_2 f_3 
+ 18 f_0^2 f_3^2 + 24 f_0 f_1 f_4 + 32 f_0^2 f_2 f_4 \right) \\
&& \times \left[ \frac{1}{2} \pi R \Lambda \left( 
\frac{1}{\sqrt{1 + 2 f_0 f_1}} 
\coth{\left( \pi R \Lambda \sqrt{1 + 2 f_0 f_1} \right)} 
+ \pi R \Lambda {\rm \,csch}^2\left( 
\pi R \Lambda \sqrt{1 + 2 f_0 f_1} \right) \right) \right] 
\frac{1}{1 + 2 f_0 f_1} \\
&& + \left( 45 f_1^2 f_2 + 120 f_0 f_1 f_2^2 + 80 f_0^2 f_3^3 
+ 54 f_0 f_1^2 f_3 + 144 f_0^2 f_1 f_2 f_3 
+ 96 f_0^3 f_2^2 f_3 \right) \\
&& \times \left[ \frac{3}{8} \pi R \Lambda \left( 
\frac{1}{\sqrt{1 + 2 f_0 f_1}} \coth{\left( \pi R \Lambda 
\sqrt{1 + 2 f_0 f_1} \right)} 
+ \pi R \Lambda {\rm \,csch}^2\left( \pi R \Lambda 
\sqrt{1 + 2 f_0 f_1} \right) \right) 
\frac{1}{{(1 + 2 f_0 f_1)}^2} \right. \\
&& ~~~~ \left. + \frac{1}{4} {(\pi R \Lambda)}^3 \coth\left( 
\pi R \Lambda \sqrt{1 + 2 f_0 f_1} \right) 
{\rm \,csch}^2\left( \pi R \Lambda \sqrt{1 + 2 f_0 f_1} \right) 
\frac{1}{{(1 + 2 f_0 f_1)}^{\frac{3}{2}}} \right] \\
&& - \left( \frac{81}{4} f_1^4 + 108 f_0 f_1^3 f_2 
+ 216 f_0^2 f_1^2 f_2^2 + 192 f_0^3 f_1 f_2^3 + 64 f_0^4 f_2^4 
\right) \\
&& \times \left[ \frac{5}{16} \pi R \Lambda \left( 
\coth\left( \pi R \Lambda \sqrt{1 + 2 f_0 f_1} \right) 
\frac{1}{\sqrt{1 + 2 f_0 f_1}} - \pi R \Lambda {\rm \,csch}^2\left( 
\pi R \Lambda \sqrt{1 + 2 f_0 f_1} \right) \right) 
\frac{1}{{(1 + 2 f_0 f_1)}^3} \right. \\
&& ~~~~ + \, \frac{1}{4} {(\pi R \Lambda)}^3 \coth\left( 
\pi R \Lambda \sqrt{1 + 2 f_0 f_1} \right) 
{\rm \,csch}^2\left( \pi R \Lambda \sqrt{1 + 2 f_0 f_1} \right) 
\frac{1}{{1 + 2 f_0 f_1}^{\frac{5}{2}}} \\
&& ~~~~ \left. + ~ {(\pi R \Lambda)}^4 \left( 
\frac{1}{12} \coth^2\left( \pi R \Lambda \sqrt{1 + 2 f_0 f_1} \right) 
{\rm \,csch}^2\left( \pi R \Lambda \sqrt{1 + 2 f_0 f_1} \right) 
+ \frac{1}{24} {\rm \,csch}^4\left( 
\pi R \Lambda \sqrt{1 + 2 f_0 f_1} \right) \right) \right. \\
&& ~~~~ \left. \times \frac{1}{{(1 + 2 f_0 f_1)}^2} \right] \\
\end{eqnarray*}

}

\end{document}